\documentstyle[12pt]{article}

\newcommand{\dslash}{\partial\!\!\!/}
\newcommand{\aslash}{A\!\!\!/}

\newcommand{\ba}{\begin{eqnarray}}
\newcommand{\ea}{\end{eqnarray}}
\newcommand{\be}{\begin{equation}}
\newcommand{\ee}{\end{equation}}

\begin{document}

\thispagestyle{empty}
%\begin{raggedleft}
%IF-UFRJ/xx/97\\
%hep-th/9707204\\
%July/97\\
%\end{raggedleft}
$\phantom{x}$\vskip 0.618cm

%\vfill

%\vfill

\begin{center}
{\huge  Topological Mass Generation as an Interference Phenomenon.}\\
[3ex]{\Large A. Ilha and C. Wotzasek}\\
[3ex]{\em Instituto de F\'\i sica\\Universidade Federal do Rio de Janeiro\\
21945, Rio de Janeiro, Brazil\\}
\end{center}

\begin{abstract}
\noindent It is proposed a new mechanism for the phenomenon of topological mass generation in three spacetime dimensions as the result of the interference of two opposite massless chiral modes. This mechanism, already used to produce the massive vectorial mode of the 2D Schwinger model, is here exploited to produce the gauge invariant massive mode of the Maxwell-Chern-Simons theory. Moreover the procedure is clearly dimensionally independent: a new chiral boson action is proposed for odd and even dimensional space-times to be used as the basic building blocks of the interference schemes.  This is a new result that extends the two-dimensional Floreanini-Jackiw action to higher dimensional spaces and is in clear contrast with the twice odd dimensional chiral form extensions.
\end{abstract}
\newpage
%%%%%%%%%%%%%%%%%%%%%%%%%%%%%%%%%%%%

\section{Introduction}

Since the advent of unification schemes based on non-abelian gauge models the quest for vectorially massive propagating modes has become phenomenologically important.  The electroweak mechanism for gauge symmetry breaking, the Higgs mechanism, is viewed by many physicists as esthetically unattractive and phenomenologically unsatisfactory.
In spite of many efforts, a realistic mechanism to give mass to gauge fields has not been found.  For some attempts, the solution was found by the addition of extra fields not required by the phenomenology, like technifermions and Higgs superparticles.  Attempts to replace the Higgs scalars by fermionic bound states have also been considered unsatisfactory.  It is interesting to compare these ideas with the so called dynamical gauge symmetry breaking mechanisms that do not require the presence of Higgs - the two dimensional Schwinger model and the three dimensional Maxwell-Chern-Simons theory (MCS). 

We propose a new interpretation for the phenomenon of dynamical mass generation as a consequence of quantum interference of massless ``chiral" modes based on the soldering formalism\cite{MS}.  The interference mechanism becomes the fundamental principle unifying the Schwinger mechanism\cite{JS} with the topological mass generation provided by the Chern-Simons term\cite{Deser:1982wh}.
An explicit realization of this phenomenon in two dimensions using the soldering formalism was found\cite{ABW}.  It was shown that fusing the massless modes of two chiral Schwinger models\cite{JR} of oppositely chiralities produces the massive vector mode present in the Schwinger model.  Here we extend the use of the soldering formalism to fuse the massive mode of the MCS out of the two massless chiral modes in (2+1) dimensions which are dimensional extensions of the 2D chiral bosons\cite{WSFJ,MWYW}.
In the next section we study the connection of these models by dimensional reduction.  The soldering mechanism for the two-dimensional case is reviewed in section 3.  Section 4 contains our main proposals: a derivation of a three dimensional chiral boson using the dual projection procedure\cite{CW} and their use in the soldering algorithm to obtain the topologically massive mode of the MCS.  We present our conclusions in section 5.

\section{Dimensional Reduction}

The description of the topologically massive mode of the MCS theory as the interference of chiral modes is signalized from its connection by dimensional reduction with the Schwinger model. This technique works by expanding the fields in normal modes corresponding to the compatified dimensions and forms the basis for the modern Kaluza-Klein theories.  To our purposes here we consider a more restricted class of reduction in which the fields are independent of the extra dimensions.  To perform the dimensional reduction of the MCS theory,

\be
{\cal L}_{MCS}= -\frac 14 F_{\mu\nu}F^{\mu\nu} + \frac m2 \epsilon^{\mu\nu\lambda}A_\mu\partial_\nu A_\lambda ,
\ee
we split the basic potential as $A_\mu = (A_a,\phi)$. The dimensional reduction is effected by assuming the potentials to be independent of $x_2$, for instance, which produces\cite{Govindarajan:1985ff}

\be
{\cal L}_{{MCS}}|^{red} = -\frac 14 F_{ab}F^{ab} +\frac 12\left(\partial_a\phi\right)^2 - m \tilde F \phi
\ee
after dropping a surface term.  Here $\tilde F= -\frac 12 \epsilon^{ab} F_{ab}$.  We recognize this Lagrangian as the quantum Schwinger model which incorporates automatically the anomaly of the axial-vector current.  This result indicates a deep connection between the topologically massive vectorial mode of the MCS theory and that of the Schwinger model.

It is useful at this point to digress on the physical meaning of the mechanism involving these phenomena. Let us examine the residues of the propagators of the MCS theory at the position of the poles.  After including a gauge fixing term in the ${\cal L}_{g.f.}= \frac 1{2\alpha}\left(\partial_\mu A^\mu\right)^2$ in the MCS theory the transverse and longitudinal sectors of the propagators become,

\ba
\Delta_T^{\mu\nu}(k) &=& \frac{-i}{k^2(k^2 -m^2)}\left[i\;m\epsilon^{\mu\lambda\nu}k_\lambda + k^2\left(\eta_{\mu\nu}-\frac{k^\mu k^\nu}{k^2}\right)\right]\nonumber\\
\Delta_L^{\mu\nu}(k) &=& \frac{i\alpha}{k^2}\left(\frac{k^\mu k^\nu}{k^2}\right) .
\ea
The longitudinal propagator has a massless pole only while the transverse propagator has both massless ($k^2=0$) and massive ($k^2-m^2=0$) poles.  Since $k^2 \geq 0$ there is no tachyon and causality is warranted. To check unitarity we couple the 2-point Green functions to external conserved currents that saturates the propagators and compute the amplitudes $\hat A_i = J^*_\mu \Delta_i^{\mu\nu} J_\nu\nonumber$, $i=L,T$. Unitarity then requires that the imaginary part of the residues of these amplitudes at the position of the poles must be nonnegative\cite{Froissart:1961ux},

\be
\mbox{Im}\, \mbox{Res}\, \hat A_i {\mid}_{\mbox{poles}} \geq 0 \;\;\; ;\; i=T,L .
\ee
To compute the amplitudes we choose $k^\mu =(m, 0, \pm m)$ for the massless pole and $k^\mu = (m, 0, 0)$ for the massive one. Then, after imposing current conservation ($k^\mu J_\mu =0$) we obtain, 

\ba
\mbox{Im}\, \mbox{Res}\, \hat A_i {\mid}_{k^2=0} &=& 0\nonumber\\
\mbox{Im}\, \mbox{Res}\, \hat A_T {\mid}_{k^2=m^2} &=& \mid J_1 + i\, J_2 \mid^2 > 0
\ea
showing that the quantum excitation associated to the long range pole is non-propagating.  On the other hand from
it is clear that the propagating excitations are indeed massive.  The connection with the dimensionally reduced Schwinger model is seen by noticing that the MCS theory contains Nielsen-Olesen string-like solutions.  The long range nature of the potentials is inferred from the equations of motion coupled to external sources, whose $\nu=0$ component is

\be
\label{gauss}
{\bf\nabla} . {\bf E} - \frac m2 B = J^0 .
\ee
Integrating over the whole plane gives,

\be
m\oint{\bf A} . {\bf {dx}} \not= 0
\ee
showing that ${\bf A }$ is a long ranged potential whose field strength is short ranged.
The Schwinger model, on the other hand, has a global axial symmetry whose current is not conserved due to the presence of the anomaly,

\be
\partial_\mu j^\mu_A=- m \tilde F .
\ee
Since in the Schwinger model the anomaly is critically responsible for giving mass to the gauge boson we conclude that the topological mass that is responsible for the long range potential is also responsible for the anomaly in the dimensionally reduced SM.

\section {Soldering and Interference in 2D}

To introduce the basic concepts of the soldering formalism let us briefly examine the soldering of two chiral Schwinger models\cite{ABW}.  The
explicit one loop calculation of the fermionic determinant, following Schwinger's point splitting method yields, in bosonized language

\begin{eqnarray}
\label{30}
W_\pm^{(0)}[\varphi] &=&-i \log \det\, (i\dslash+e\aslash_\pm)\nonumber\\
&=&{1\over{4\pi}}\int d^2x\,\left(\partial_+
\varphi \partial_-\varphi +2 \, e\,A_\pm\partial_-\varphi + a_\pm\, 
e^2\, A_+ A_-\right).
\end{eqnarray}
Note that the regularization ambiguity is manifested through the Jackiw-Rajaraman parameters $a_\pm$ which are arbitrary except that $a_{\pm} \geq 1$ to avoid tachyonic excitations.
The spectrum of these models has been carefully studied and shown to fall in three distinct classes characterized according to the number of second-class constraints present in the Hamiltonian approach.  The original study\cite{JR} $(a>1)$ has disclosed the presence of a massive and a massless excitations while the four constraints class $(a=1)$ studied in \cite{RR} has only displayed a massless particle.  A new class of solutions with three constraints was latter studied in \cite{PM} and \cite{Abreu:2000dv}.  The spectrum was shown to be analogous to the two constraint class but the massless excitation is chiral.

After implementing the soldering, one finds a Polyakov-Weigman type effective action containing a current-current interference term\cite{PW},  

\be
W\left[ \Phi(\varphi, \rho\right)] = W_+^{(0)}\left[ \varphi\right] + W_-^{(0)}\left[ \rho\right] + \frac 12\int d^2x 
\left[J\left(\varphi ,\rho\right)\right]^2 .
\ee
The Noether current in the interference piece is

\ba
J\left(\varphi ,\rho\right) &=& J_+\left(\varphi\right) + J_-\left(\rho\right)\nonumber\\
J_{\pm}\left(\varphi\right) &=& \frac 1{2\pi} \left(\partial_{\pm}\varphi  + e A_{\pm}\right) .
\ea
The effective soldered action now reads

\begin{equation}
\label{110}
W[\Phi]={1\over {4\pi}}\int d^2x\:\Big{\{}\Big{(}\partial_+
\Phi\partial_-\Phi + 2\,e\, A_+\partial_-\Phi - 2\,e\, A_-
\partial_+\Phi\Big{)} +(a_+ + a_- -2)\,e^2\,A_+\,A_-\Big{\}}
\end{equation}
where $\Phi=\varphi - \rho$ is the collective field, invariant under the soldering transformation.  As discussed in \cite{ABW} this action reduces to the usual gauge invariant Schwinger model for the case $a_+ = a_- =1$ which corresponds to the four constraints regularization class and massless spectrum.  This shows that the massive mode in the Schwinger model is the result of the interference between right and left chiral modes.

\section {3D Chiral Bosons and Soldering}

Let us next discuss the three dimensional ``chiral" modes.
We define chiral modes, in a dimensionally independent way, as the half degree of freedom of a massless scalar field.  To construct it we reduce the phase space by imposing a chiral like constraint in the first-order action of the model.  The resulting chiral mode is described by an action similar in form to the well known Floreanini-Jackiw chiral boson in (1+1) dimension. 
This result {\it per se} is already quite surprising since there exists a strong belief that chiral bosons only exist in spacetimes of (twice odd) even dimensions (D=4k-2 ; $k\in Z_+$) and certainly not in odd dimensional spacetimes\cite{many}.
To explicitly construct the chiral modes we write the massless scalar action in its first-order form

\be
S[\phi] = \int dx\, \left[\pi \dot \phi - \frac 12 \pi^2 -\frac 12 \phi\left(-\nabla^2\right)\phi\right]
\ee
and impose the ``chiral" constraint.  A simple inspection will show that such restriction can not be acchieved in functional space.  Indeed if a restriction like $\phi\to \phi_1$ and $\pi\to\pm\phi_2'$, with $\phi' =\sqrt{-\nabla^2}\,\phi$, is implemented then the resulting action acquires the Schwarz-Sen structure\cite{Schwarz:1994vs} and the phase space does not get reduced.  To find the proper chiral restriction over the phase space we follow the procedure already depicted in \cite{CW} called dual projection.  This technique discloses a four dimensional phase space for the Fourier modes.  In this sense we introduce a two-dimensional basis $\{e_a({\bf k},{\bf x})\, ;\, a=1,2\}$ with $({\bf k},{\bf x})$ being a pair of conjugate variables. The vectors spanning the Fourier space satisfy an orthonormalization condition as,

\be
\label{dich10}
\int d{\bf x} \, e_a({\bf k},{\bf x})\, e_b({\bf k}' ,{\bf x}) = \delta_{ab}\, \delta({\bf k}-{\bf k}')
\ee
and are chosen to be eigenvectors of the Laplacian operator,

\be
\label{dich20}
\nabla^2 \, e_a({\bf k},{\bf x}) = - \omega^2({\bf k})\, e_a({\bf k},{\bf x}).
\ee
Using this basis to represent our elementary fields,

\ba
\label{dich30}
\phi({\bf x}, t) &=& \int d{\bf k}\, q_a({\bf k},t) e_a({\bf k},{\bf x}) \nonumber\\
\pi({\bf x},t) &=& \int d{\bf k}\, p_a({\bf k},t) \, e_a({\bf k},{\bf x}),
\ea
with $q_a$ and $p_a$ being the expansion coefficients, the Lagrangian for the scalar field is reduced to that of a two-dimensional harmonic oscillator for each mode,

\be
\label{dich40}
L= \int dk\,\left(p_a \dot q_a - \frac 12 p_a p_a  - \frac 12 \omega^2 q_a q_a\right).
\ee
The basic idea of the dual projection is to impose the chiral constraint in the Fourier space as,

\be
\label{dich50}
p_a({\bf k},t) = \pm \,\omega({\bf k}) \,\epsilon_{ab}\, q_b({\bf k},t).
\ee
This procedure reduces the four dimensional phase space and the resulting Lagrangian, 

\be
\label{dich60}
L_\pm = \int d{\bf k}\, \omega \left[\pm\dot q_a^{(\pm)}\epsilon_{ab}\, q_b^{(\pm)} - \omega \, q_a^{(\pm)} q_a^{(\pm)}\right],
\ee 
represents a chiral field with each mode describing a chiral oscillator.  For the two-dimensional spacetime this representation for the chiral scalar has been proposed by Bazeia\cite{Bazeia:1991he} through the substitution

\ba
\label{dich70}
\partial_t \phi &\to &\dot q_a\nonumber\\
\partial_x \phi &\to& \omega\epsilon_{ab} q_b .
\ea
Clearly the action (\ref{dich60}) reduces to Floreanini-Jackiw action in this dimension but the dual projection procedure extends this concept to any dimension.

Once we have the at our disposal the chiral actions representing chiral scalar degrees of freedom of opposite chirality the stage is set for the soldering formalism.  As discussed above\cite{ABW} we want to construct an effective theory in the form,

\be
\label{eff}
L_{eff} = L_{+}^{(0)}(q_a^+) + L_{-}^{(0)}(q_a^-) + L_{int}(q_a^+, q_a^- )
\ee
with the chiral components $q_a^\pm$ being considered as independent variables and the interference term having the general form of current-current coupling\cite{rabin}

\be
L_{int} = \frac 12 \int d{\bf k} \,  {\cal J}(q_a^+,q_a^-) . {\cal J}(q_a^+,q_a^-)
\ee
but with the soldering current being a separable function of both chiral fields as

\be
{\cal J}(q_a^+,q_a^-) = J_+ (q_a^+) + J_-(q_a^-) .
\ee
For the case at hand we find

\be
J_a^{(\pm)}= \pm \, e\, \epsilon_{ab}\, q_a^{(\pm)}
\ee
where $e$ is a coupling constant.  It is interesting to notice that the Noether current has the same form as in the Chiral Schwinger model and reduces to it by the dimensional mechanism.

Next we redefine the fields in terms of the symmetric and antisymmetric combinations of $q_a^\pm$ and introduce a suggestive electromagnetic notation as,

\be
\label{dich80}
\pi_i({\bf x},t) = \int d{\bf k}\,\left[M_{ijab}^{(+)} \, q_a^{(+)}({\bf k},t) + M_{ijab}^{(-)}\, q_a^{(-)}({\bf k},t)\right]\hat\partial_j \, e_b({\bf k},{\bf x})
\ee
where 

\ba
\hat\partial_j &=& \frac{\partial_j}{\sqrt{-\nabla^2}}\nonumber\\
M_{ijab}^{\pm} &=& \omega({\bf k}) \epsilon_{ij}\delta_{ab} \pm e \, \delta_{ij} \epsilon_{ab}
\ea
and

\be
\label{dich90}
A_i({\bf x},t) = \int d{\bf k} \, \epsilon_{ab}\epsilon_{ij} \left[q_a^{(+)}({\bf k},t) - q_a^{(-)}({\bf k},t)\right] \hat\partial_je_b({\bf k},{\bf x}) + \partial \psi({\bf x},t) .
\ee
Here $\pi_i$ and $A_i$ represent the conjugate pair of electromagnetic variables and $\psi$ is an arbitrary function reflecting the longitudinal ambiguity of the electromagnetic potential.  Notice that, as defined by (\ref{dich90}), the (scalar) magnetic field is,

\ba
\label{dich100}
B({\bf x},t) &=& - \epsilon_{ij} \partial_i A_j({\bf x},t)\nonumber\\
&=& -\int d{\bf k} \, \omega({\bf k})\,  q_a({\bf k},t)\, e_a({\bf k},{\bf x})
\ea
and satisfy the MCS constraint (\ref{gauss}) for the free case

\be
\label{dich110}
G={\bf \nabla}.{\bf \pi} - e\, B .
\ee
This constraint is next incorporate into the theory via a Lagrange multiplier, call it $A_0$.   Bringing these definitions into the effective action (\ref{eff}) and making use of the identity

\be
\label{dich120}
-\delta_{ij} = \hat\partial_i\, \hat\partial_j + \epsilon_{ik}\epsilon_{jm}\,\hat\partial_k\,\hat\partial_m
\ee
we obtain,

\be
\label{dich130}
S_{eff}= \int d^3 x\,\left[{\bf \pi}.{\bf \dot A} - \frac 12 {\bf \pi}.{\bf \pi} - \frac 12 B^2
-\frac {e^2}{2} {\bf A} .{\bf A} - e {\bf A}. \epsilon .{\bf \pi} +A_0 G\right]
\ee
which is the first-order action for the Maxwell-Chern-Simon theory, with coupling constant playing the role of the mass of the vectorial mode, $e=m/2$.  Observe that by solving for the $\pi_i$ field, the resulting second-order action is,

\be
{\cal L}_{eff} = {\cal L}_{MCS} (A_k,A_0)
\ee
and depends only on ``collective" fields that are functions of the anti-symmetric combination of the chiral scalars, i.e., it depends only on the invariant combination $Q_a = q_a^+ - q_a^-$ of the chiral variables.  This is similar to the behavior of the soldered action of chiral scalar yielding the vectorial Schwinger model in 2D.

\section{Conclusions}

In this paper we have proposed a new interpretation for the vectorial massive mode of the topologically massive Maxwell-Chern-Simons theory as an interference phenomenon.  This mechanism would parallel a similar phenomenon in two-dimensional field theory known as Schwinger mechanism that results from interference between right and left massless chiral scalars.
To this end we have proposed a new chiral boson theory defined in any dimensions, both odd and even.  This is a new result that extends the notion of D=2 chiral scalars defined in the Floreanini-Jackiw theory. The chiral action results from a constraint imposed over the Fourier modes of the ordinary first-order scalar action as a twisting between the canonical components of the fields.  The corresponding action in functional space results being of nonlocal character.  For the D=3 case, the soldering of these chiral components leads directly to the topologically massive Maxwell-Chern-Simos theory. This result unifies both phenomena of mass generation as a consequence of chiral interference and confirms the connection among the models already given by dimensional reduction.

\vspace{1cm}

\noindent {\bf Acknowledgment:} This work is supported in part by
CNPq, CAPES, FINEP and FUJB (Brazilian Research Agencies).
 
%\newpage


\begin{thebibliography}{30}
\bibitem{MS}M. Stone, Phys. Rev. Lett. 63 (1989) 731; Nucl. Phys. B 327 (1989) 399; Report No. ILL-23/89 (unpublished); D. Depireux, S. J. Gates, Jr. and Q-Han Park,  Phys. Lett. 224B (1989) 364; E.~Witten,
%``On Holomorphic factorization of WZW and coset models,''
Commun.\ Math.\ Phys.\  {\bf 144}, 189 (1992).
%%CITATION = CMPHA,144,189;%%
%\href{\wwwspires?j=CMPHA%2c144%2c189}{SPIRES}

\bibitem{JS}   J.Schwinger, Phys. Rev. 128 (1962) 2425.

\bibitem{Deser:1982wh}
S.~Deser, R.~Jackiw and S.~Templeton,
%``Topologically Massive Gauge Theories,''
Annals Phys.\  {\bf 140}, 372 (1982).
%%CITATION = APNYA,140,372;%%
%\href{\wwwspires?j=APNYA%2c140%2c372}{SPIRES}

\bibitem{ABW}E. M. C. Abreu, R. Banerjee and C. Wotzasek, Nucl. Phys. B 509(1998)519; R. Banerjee and C. Wotzasek, Nucl. Phys. B 527 (1998) 402; R. Amorim, A. Das, and C. Wotzasek, Phys. Rev. D 53 (1996) 5810.

\bibitem{JR}R. Jackiw and R. Rajaraman, Phys. Rev. Lett. 54 (1985) 1219.

\bibitem{WSFJ}W. Siegel, Nucl. Phys. B 238(1984)307; R. Floreanini and R. Jackiw, Phys. Rev. Lett. 59(1987)1873.

\bibitem{MWYW}B. McClain, Y. S. Wu and F. Yu, Nucl. Phys. B 343(1990)689;
C. Wotzasek, Phys. Rev. Lett. 66(1991)129;

\bibitem{CW}C. Wotzasek, Phys. Rev. D57(1998) 4990.

\bibitem{Govindarajan:1985ff}
T.~R.~Govindarajan, S.~D.~Rindani and M.~Sivakumar,
%``Dimensional Reduction And Theories With Massive Gauge Fields,''
Phys.\ Rev.\  {\bf D32}, 454 (1985); 

\bibitem{Froissart:1961ux} 
M.~Froissart,
%``Asymptotic Behavior And Subtractions In The Mandelstam Representation,''
Phys.\ Rev.\  {\bf 123} (1961) 1053; A.~Martin,
%``Unitarity And High-Energy Behavior Of Scattering Amplitudes,''
Phys.\ Rev.\  {\bf 129}, 1432 (1963); O.~M.~Del Cima,
%``Probing the Froissart bound for models with charged vector fields in D = 3,''
Mod.\ Phys.\ Lett.\  {\bf A9}, 1695 (1994).

\bibitem{RR}   R. Rajaraman, Phys. Lett. B 154 (1985) 305.

\bibitem{PM}   P. Mitra, Phys. Lett. B 284 (1992) 23.

\bibitem{Abreu:2000dv}
E.~M.~Abreu, A.~Ilha, C.~Neves and C.~Wotzasek,
%``Interference phenomenon for the Faddeevian regularization of 2D chiral fermionic determinants,''
Phys.\ Rev.\  {\bf D61}, 025014 (2000).

\bibitem {PW} A. Polyakov and P. Weigman, Phys. Lett. B131 (1983) 121

\bibitem{many} R. Manvelyan, R. Mkrtchyan and H.J.W. Muller-Kirsten, Phys. Lett. B453 (1999) 258; W.T. Kim, Y.W. Kim and Y.J. Park, Jour. Phys. A32 (1999) 2461; N. Berkovits and C. Hull, J. High Energy Phys. 2 (1998) U195.


\bibitem{Schwarz:1994vs}
J.~H.~Schwarz and A.~Sen,
%``Duality symmetric actions,''
Nucl.\ Phys.\  {\bf B411}, 35 (1994)

\bibitem{Bazeia:1991he}
D.~Bazeia,
%``A Quantum mechanical anomaly reexamined,''
Mod.\ Phys.\ Lett.\  {\bf A6}, 1147 (1991).


\bibitem{rabin} R.Banerjee and B. Chakraborty, Jour. Phys. A32 (1999) 4441; R. Banerjee and S. Kumar, Phys. Rev. D60:085005, 1999.

\end{thebibliography}
\end{document}